\documentclass[aps,pre,twocolumn,showpacs,superscriptaddress,superscriptaddress,longbibliography]{revtex4-1}  % for review and submission
\usepackage{graphicx,color}  % needed for figures
\usepackage{dcolumn}   % needed for some tables
\usepackage{bm}        % for math
\usepackage{amssymb}   % for math

\usepackage{ulem}
\usepackage[english]{babel} 
\usepackage{amssymb,amsfonts,amsmath}
\usepackage{graphicx,bm}
\usepackage{color}
\usepackage{epsfig,dsfont,amssymb,amsmath,amsthm,amsfonts,amsbsy,mathrsfs}
\usepackage{color}
\usepackage{xcolor}
\usepackage{tikz}

\definecolor{red}{rgb}{1,0,0}
\definecolor{blue}{rgb}{0,0,1}
\definecolor{black}{rgb}{0,0,0}
 
\usepackage{amsmath}
\begin{document}

\title{Microscopic processes controlling the Herschel-Bulkley exponent}
\author{Jie Lin}
\affiliation{School of Engineering and Applied Sciences, Harvard University, Cambridge, Massachusetts 02138, USA}
\author{Matthieu Wyart}
\affiliation{Institute of Theoretical Physics, Ecole Polytechnique Federale de Lausanne (EPFL), CH-1015 Lausanne, Switzerland}

\date{\today}
\begin{abstract}
The flow curve of various yield stress materials is singular as the strain rate vanishes, and can be characterized by the so-called Herschel-Bulkley exponent $n=1/\beta$. A mean-field approximation due to Hebraud and Lequeux (HL) assumes mechanical noise to be Gaussian, and leads to  $\beta=2$ in rather good agreement with observations. Here we prove that  the improved mean-field model where the mechanical noise has fat tails instead leads to $\beta=1$ with logarithmic correction. This result supports that HL is not a suitable explanation for the value of $\beta$, which is instead significantly affected by finite dimensional effects. From considerations on elasto-plastic models and on the limitation of speed at which avalanches of plasticity can propagate, we argue that $\beta=1+1/(d-d_f)$ where $d_f$ is the fractal dimension of avalanches and $d$ the spatial dimension. Measurements of $d_f$ then supports that $\beta\approx 2.1$ and $\beta\approx 1.7$ in two and three dimensions respectively. We discuss theoretical arguments leading to approximations of $\beta$ in finite dimensions.

%Using a simple approximation on the avalanche statistics, we obtain  $\beta\approx 1+(1+\theta)/d(1-\theta)$ where $d$ is the spatial dimension and $\theta$ the pseudo gap exponent characterizing the density of shear transformations close to yielding amenable to  calculations, leading to an approximate analytic result $\beta\approx 1.6$. 
\end{abstract}

\maketitle

\section{Introduction}
The flow curves of various amorphous materials including emulsions and foams is singular near the yielding transition, and can be written as $\dot{\gamma}\sim (\sigma-\sigma_c)^{\beta}$ where $\dot{\gamma}$ is the strain rate and $\sigma_c$ the yield stress \cite{Herschel26}. The inverse of $\beta$,   $n=1/\beta$,  is the so-called Herschel-Bulkley (HB) exponent. 
Reported values of $\beta$ differ, but are often in the range $[1.4,2.5]$ \cite{Salerno12,Chaudhuri12,Mobius10,Becu06}.
On reason for these discrepancies is that $\beta$ is sometimes extracted from data that includes large stresses $(\sigma-\sigma_c)\sim \sigma_c$, where non-linear phenomena can affect the dissipation mechanism (e.g. changing the film thickness between bubbles in a foam \cite{Cloitre2003}). Instead here we focus on the vicinity of the yielding transition $(\sigma-\sigma_c)\ll \sigma_c$, where the Herschel-Bulkley exponent reveals the collective nature of the dynamics near a dynamical critical point. We know since Argon that plastic flow in these materials  can be decomposed into local rearrangements involving a few particles \cite{Argon79}, called shear transformations. One shear transformation affects the stress around it, which can in turn trigger new shear transformations \cite{Lemaitre09, Picard2005,Vandembroucq2011}. In the solid phase, plasticity thus occurs by burst of shear transformations or avalanches \cite{Salerno12,Talamali11,Jagla15} which are system spanning for very slow drive in the entire solid phase $\sigma<\sigma_c$ \cite{Lin15a}.  Scaling relations can be derived relating $\beta$ to other exponents describing the avalanche fractal dimension and their duration \cite{Lin14}. However,  predicting the value of $\beta$ remains a challenge.  The most popular theoretical approach is arguably the  model of Hebraud and Lequeux, where the mechanical noise generated by the relaxation of shear transformations  is treated in a mean-field manner (neglecting spatial correlations) and assumed to be  Gaussian  \cite{Hebraud98,Agoritsas15,Agoritsas17}, leading to a value $\beta=2$, which appears at first sight to be a reasonable number to explain experiments.

Recently we have shown that the assumption of Gaussian mechanical noise leads to qualitatively wrong predictions for important structural properties. In particular, a central aspect of these materials near their yield stress is the density of shear transformations about to become unstable. This property can be computed within good approximation in a mean-field theory where the mechanical noise has fat tails \cite{Lin16}, which results from the long-range interaction of shear transformations \cite{Lemaitre07}. Here we show that in this more accurate mean-field approach, $\beta=1$ (with a logarithmic correction), as we confirm numerically. This result supports that finite dimensional effects must be included to obtain reasonable description of the flow curve. In the second part of this manuscript, we argue based on the limitations at which information can propagate in elastic materials that $\beta=1+1/(d-d_f)$ where $d_f$ is the fractal dimension of avalanches and $d$ the spatial dimension. We use this relation to extract estimates of $\beta$ from elasto-plastic models (cellular automata). We obtain $\beta\approx 2.1$ for $d=2$ and $\beta\approx 1.7$ for $d=3$. Using a simple approximation on the avalanche statistics that should improve with dimensions and is already good for $d=3$, we obtain an estimate of $d_f$ leading to $\beta\approx 1+(1+\theta)/d(1-\theta)$ where $d$ is the spatial dimension and $\theta$ the pseudo gap exponent characterizing the density of shear transformations close to yielding. $\theta>0$ is imposed by stability \cite{Lin14a}, and its value can be computed within a mean-field  calculation \cite{Lin16}, leading to an approximate analytic result $\beta\approx 1.6$ for $d=3$.

% We argue that in large dimensions, $\beta\approx 1+(1+\theta)/d(1-\theta)$ where $\theta$ is the pseudo-gap exponent characterizing the density of shear transformations. This expression already leads to a good estimate in three dimensions. 
%
%This supports the larger values of $\beta$ reported in experiments and simulations are in fact due to finite dimensional effects.  and we derive an approximate expression of $\beta$ based on the elasto-plastic model in the end. Section \ref{meanfield} introduces the mean-field models in both discrete and continuous version. Theoretical analyses on the distribution of local stabilites and the HB exponent near the critical stress are in Sec. \ref{theory}.  Predictions are tested numerically in Sec. \ref{numerics} and the implications of our mean field model on real materials are discussed in Sec. \ref{finitedimension}. 

\section{\label{meanfield}Elasto-plastic models }
\subsection{General framework} 
In  elasto-plastic models \cite{Picard2005,Talamali11,Nicolas14} amorphous solids are discretized into $N$ mesoscopic sites  on a square or cubic lattice. Each lattice site represents a few particles and carries a local shear stress $\sigma_i$, so that the total shear stress applied on the system is $\sigma=\langle \sigma_i\rangle$ where the average is taken on all sites.   Each site is characterized by a local yield stress $\sigma_i^{th}$ \cite{Talamali11,Picard2005,Martens11,Liu16}, assumed here to be unity (adding spatial disorder on this quantity does not change our conclusions). When $\sigma_i>1$, the probability per unit time for this site to become plastic is not zero and is given by $1/\tau_c$ \cite{Princen83}. The stress on site $i$ then goes to zero, and is redistributed to other sites immediately as:
\begin{equation}
\sigma_j(l+1)=\sigma_j(l)+\mathcal{G}_{ji}(\vec{r}_i-\vec{r}_j) \sigma_i (l),
\end{equation}
where $\mathcal{G}_{ji}(\vec{r}_i-\vec{r}_j)$ is the interaction kernel, and the integer $l$ numbers plastic events in chronological order. Note that in such models,  the stress changes  instantaneously in the entire system (see more on that below). 
 If the shear stress is fixed, one must have $\sum_{j\neq i} {\cal G}_{ij}=-{\cal G}_{ii}=1$. In $d$ dimensions, the elastic kernel follows $\mathcal{G}(\vec{r})=f(\hat{n})/r^{d}$  \cite{Picard04}, where $f(\hat{n})$ is the angle-dependent factor which can be positive or negative depending  on the direction  $\hat{n}$, the unit vector between sites $i$ and $j$. 
 
 \subsection{Discrete mean-field model} 
 In the mean field model, spatial correlations are destroyed by drawing at each event $l$ a random permutation ${\cal P}_l$  of the $N-1$ sites which did not become plastic, so that:
 \begin{equation}
\sigma_j(l+1)=\sigma_j(l)+\mathcal{G}_{{\cal P}_l(j)i} \sigma_i (l),
\end{equation}
In such a model the only important aspect of $\mathcal{G}$ is its distribution $\rho(\delta \sigma)$ characterizing the probability that a site $j$ gets a change of stress of amplitude $\delta \sigma$. A straightforward integration on space of $\mathcal{G}(\vec{r})$ gives \cite{Lemaitre07}:
\begin{equation}
\label{rho}
\rho(\delta \sigma)= A |\delta \sigma|^{-\mu-1}/N,
\end{equation}
where $\mu=1$ and $A$ is a constant (which can be expressed in terms of the function $f$, and thus depends on $d$). This distribution has a lower cut-off $\delta\sigma_{min}\sim N^{-1/\mu}$ (corresponding to stress perturbation across the system extension) and  upper cut-off is $\delta\sigma_{max}\sim1$ (corresponding to adjacent sites in the original problem) \cite{Lin16}. 
This result can be readily  extended  to other interaction kernels. For a long-ranged interaction which decays as $|\mathcal{G}(r)|\sim r^{-\alpha}$, one finds that Eq. (\ref{rho}) still applies with $\mu=d/\alpha$.

 The dynamical rule following the instability of site $i$ at time $l$ becomes:
\begin{align}\label{dynamics}
\sigma_i(l+1)&= 0,\nonumber \\ 
\sigma_j(l+1)-\sigma_j(l)&= \frac{\sigma_i(l)}{N-1} +\delta \sigma_j ,
\end{align}
where $\delta\sigma_j$ is chosen randomly from the distribution Eq. (\ref{rho}). Each event is assumed to relax the strain by $1/N$, so that the total plastic strain  of the system is simply $\gamma(l)=l/N$.

\subsection{Continuous formulation} Given the above discrete dynamical model, a Fokker-Planck equation for the local distances to yielding $x_i=1-\sigma_i$ can be obtained by taking the limit $N\rightarrow \infty$ \cite{Lin16}. One obtains:
\begin{align}
&\frac{\partial P(x,t)}{\partial t}=\dot{\gamma}\{v\frac{\partial P}{\partial x} + \int_{-\infty}^{\infty} A\frac{P(y)-P(x)}{|x-y|^{\mu+1}} dy +\delta (x-1)\}\nonumber \\
&-\Theta(-x)\frac{P(x)}{\tau_c},\label{fp1}
\end{align}
Here $\Theta(x)$ is the Heaviside step function. The delta function describes the sites returning to $x=1$ right after an instability. The integral term represents the anomalous diffusion process (a Levy flight in that case) caused by the broad mechanical noise. The strain rate $\dot{\gamma}$ corresponds to the plastic activity. For a finite system size $N$, the number of plastic events during a time interval $\delta t$ is $N\dot{\gamma}\delta t$\footnote[1]{In the original version of the H\'ebraud-Lequeux\cite{Hebraud98} model, the strain rate is defined following our notation as $v\dot{\gamma}$. As far as scaling is concerned, this is equivalent to our treatment because  $|v-v_c|<<v_c$ close to yielding.}. A critical shear stress $\sigma_c$ is needed to maintain a finite strain rate $\dot{\gamma}>0$ as indicated in Eq. (\ref{fp1}). After integrating the right side of Eq. (\ref{fp1}) from $-\infty$ to $\infty$, one finds that it must follow the self-consistent relationship:
\begin{equation}
\label{11}
\dot{\gamma}=\frac{1}{\tau_c}\int_{-\infty}^{0}P(x) dx.
\end{equation}
In what follows we shall choose $\tau_c=1$. Finally, $v$ is the drift term, caused by the second  term in the right side of Eq. (\ref{dynamics}). It can be thought as a Lagrange multiplier which fixes the total applied stress $\sigma$, and is fixed through the relationship:
\begin{equation}
\label{sigma}
\sigma=\int P(x) (1-x) dx
\end{equation}
Thus to obtain the flow curve, one must solve for the stationary solution of Eq. (\ref{fp1}) for any parameters $\dot\gamma$ and $v$, and then use Eq. (\ref{11}) to fix the relationship $v(\dot\gamma)$. Finally Eq. (\ref{sigma}) can be used to fix  $\sigma(\dot{\gamma})$. 

\begin{figure}[bht!]
 \center \includegraphics[width=0.48\textwidth]{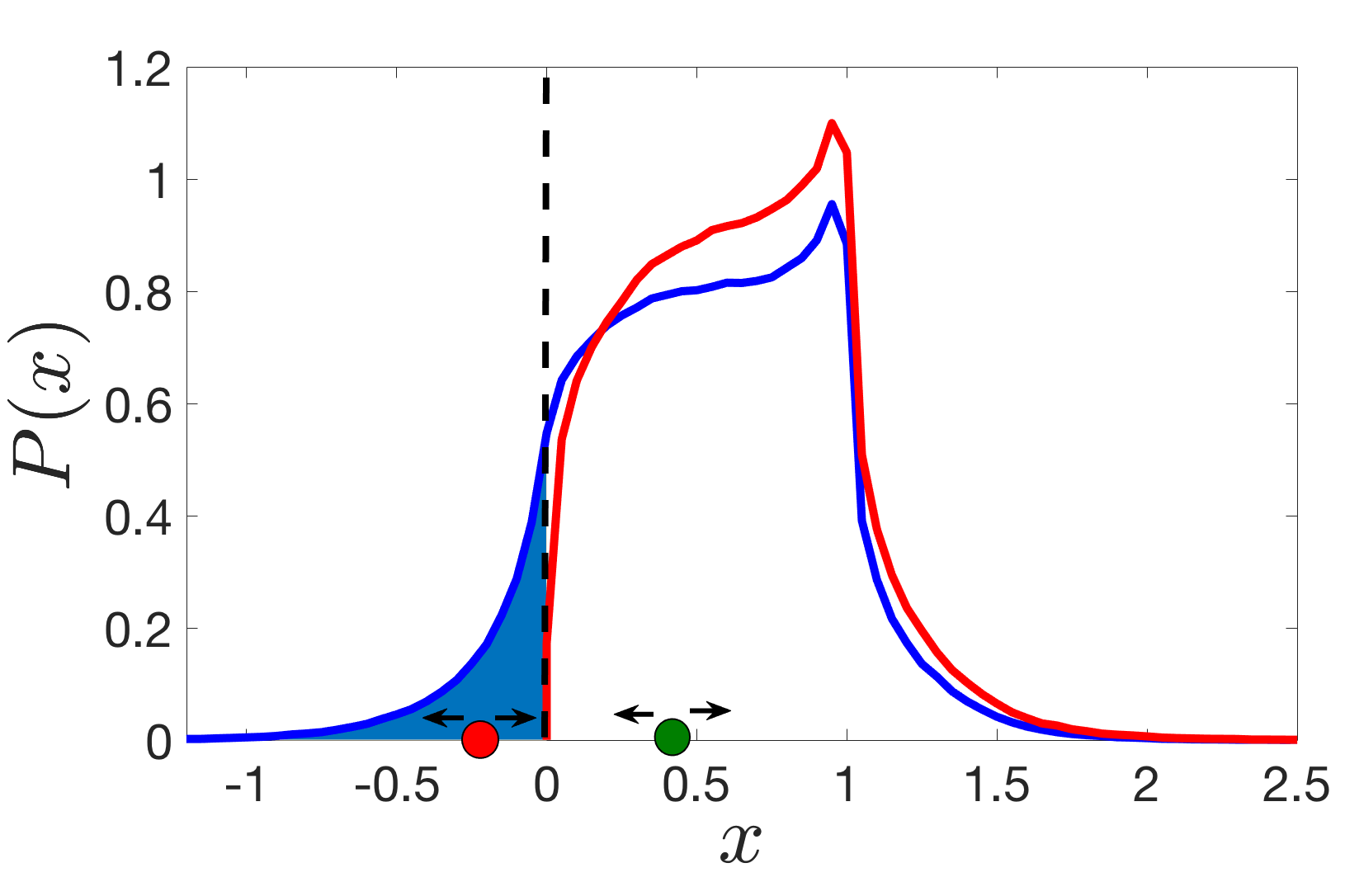}
\caption{Distribution $P(x)$ of local stabilities $x$ at $\sigma>\sigma_c$ (blue) and at $\sigma_c$ (red). The area in the negative $x$ corresponds to the strain rate $\dot{\gamma}$. The green (red) circle represents  a (un)stable site, which implements random long jumps and drifts towards the negative, unstable direction.}\label{cartoon1}
\end{figure}

We illustrate the dynamics of local stabilities $x$ of the mean field model in Fig. \ref{cartoon1}, where sketches of distributions $P(x)$ are shown for $\sigma>\sigma_c$ and $\sigma=\sigma_c$.

\section{\label{theory}Theoretical Analysis}

\subsection{  Critical density of stability $P_c(x)$}
At the critical stress $\sigma_c$, the strain rate is zero and all sites are stable, therefore $P_c(x)=0$ for $x<0$. We have shown previously that at that point, $P_c(x)$ is singular and displays a pseudo gap at $x=0$\cite{Lin16}:
 \begin{equation}
 P_c(x)\sim x^{\theta},\label{6}
 \end{equation}
 where the $\theta$ exponent depends on the interaction index $\mu$. We summarize the results here,
 \begin{align}
 \theta & = \mu/2    &1<\mu<2 , \\
 \theta & = \frac{1}{\pi} \arctan{  (\frac{\pi A}{v_c})  }  &\mu=1 ,\\
 \theta &=0    &\mu<1.  \label{7}
 \end{align}
For the physical case $\mu=1$, the pseudo gap exponent depends on the amplitude $A$ of elastic kernel and the drift $v_c$ at $\sigma_c$. 
For our purpose it is useful to obtain a closed equation for $P_c(x)$ by considering the stationary state of Eq. (\ref{fp1}). Dividing this equation by the strain rate, one obtains a term $\Theta(-x)P(x)/\dot{\gamma}$ corresponding to the normalized probability that a site yields at position $x$. 
In the limit $\dot\gamma\rightarrow 0$, the mechanical noise and the drift per unit time in Eq. (\ref{fp1}) vanishes, and sites yields at the position where they first became unstable. For $1\leq \mu<2$,  $P_c(x=0)$: all the sites becoming unstable do so by a  jump from a stable $y>0$ configuration toward an unstable one with $x<0$,  which occurs with a probability $ \int_{0}^{\infty} A\frac{P_c(y)}{|x-y|^{\mu+1}}dy$. For $\mu<1$, $P_c(x=0)>0$ and a finite fraction of the sites become unstable by drifting and hitting $x=0$, which occurs with a probability $v_c P_c(0)$. Thus we obtain:
\begin{align}
&v_c \frac{dP_c}{dx}+\int_{0}^{\infty}A\frac{P_c(y)-P_c(x)}{|x-y|^{\mu+1}}dy+\delta(x-1) \nonumber \\
&-\Theta(-x) \int_{0}^{\infty} A\frac{P_c(y)}{|x-y|^{\mu+1}}dy -v_c P_c(0) \delta(x=0)=0.\label{fp2}
\end{align}
The above equation is valid for any $0<\mu<2$. Mathematically, the second to last term  ensures that $P_c(x)=0$ is solution for $x<0$. When $\mu<1$, the last term  plays the same role by balancing the first term at $x=0$.

\subsection{Perturbation around $P_c(x)$}
We consider a small perturbation near the yielding transition, $\sigma=\sigma_c+\delta\sigma$, $v=v_c+\delta v$, and $P(x)=P_c(x)+\delta P(x)$. Using Eq. (\ref{fp1}, \ref{fp2}), we find:
\begin{align}
&v_c \frac{d \delta P(x)}{dx}+\int_{-\infty}^{\infty} A\frac{\delta P(y)-\delta P(x)}{|x-y|^{\mu+1}}dy-\Theta(-x)\frac{\delta P(x)}{\dot{\gamma}}=\nonumber\\
&-S_1-S_2.\label{Deltax}
\end{align}
This equation for $\delta P(x)$ corresponds to the Fokker-Plank equation of a biased Levy-flight motion with a constant yielding rate for $x<0$ (left hand side term) in the presence of two sources terms $S_1$ and $S_2$,
\begin{align}
S_1&=\delta v\frac{dP_c(x)}{dx}, \label{s1}\\
S_2&=\Theta(-x)\int_{0}^{\infty} A\frac{P_c(y)}{|x-y|^{\mu+1}}dy+v_c P_c(0) \delta(x=0). \label{s2}
\end{align}
We may thus decompose $\delta P(x)$ in two contributions corresponding to each source term: $\delta P(x)=\Delta_1(x) +\Delta_2(x)$. 
Conservation of probability implies that:
\begin{equation}
\int \Delta_1(x) dx =- \int \Delta_2(x) dx \label{10}
\end{equation}
From the definition of the stress one gets the following decomposition:
\begin{align}
&\sigma-\sigma_c=\int (1-x) \Delta_1(x) dx + \int (1-x) \Delta_2(x) dx \nonumber \\&= - \int x \Delta_1(x) dx - \int x \Delta_2(x) dx\equiv \delta \sigma_1 +\delta \sigma_2\label{8}
\end{align}
%They respectively satisfy the following equations
%
%\begin{align}
%\dot{\Delta}_1(x)=&\dot{\gamma}\{  S_1+v_c \frac{d \Delta_1(x)}{dx}\nonumber
%\\&+\int_{-\infty}^{\infty} A\frac{\Delta_1(y)-\Delta_1(x)}{|x-y|^{\mu+1}}dy\}-\Theta(-x)\Delta_1(x),\label{delta1}
%\end{align}
%
%\begin{align}
%\dot{\Delta}_2(x)=&\dot{\gamma}\{S_2+ v_c \frac{d\Delta_2(x)}{dx}  \nonumber
%\\&+ \int_{-\infty}^{\infty} A\frac{\Delta_2(y)-\Delta_2(x)}{|y-x|^{\mu+1}}dy\}-\Theta(-x)\Delta_2(x),\label{delta2}
%\end{align}
%where $C_1$ is a constant of $O(1)$. 
Note that the source term $S_1$  is non-zero only for $x>0$.  In that case the asymptotic solution for $\Delta_1$ as $\dot\gamma\rightarrow 0$ is solution of:
\begin{align}
&v_c \frac{d \Delta_1}{dx}+\int_{-\infty}^{\infty} A\frac{\Delta_1(y)-\Delta_1(x)}{|x-y|^{\mu+1}}dy =-S_1\nonumber \\
&\hbox{ with  }\Delta_1(x)= 0  \hbox{ for }  x<0 \label{Deltax}
\end{align}
Because $P_c(x)$ varies on the scale $x\sim1$, so does $S_1$ and $\Delta_1$, which turns out to be negative in average due to Eq. (\ref{10}) and the positivity of $\Delta_2$ (see below). We must then have asymptotically $ \int  x \Delta_1(x) dx/ \int  \Delta_1(x) dx \equiv C_1(\dot\gamma) \rightarrow C_1\neq 0$ as $\dot\gamma\rightarrow 0$. In this limit we write:
\begin{align}
\delta \sigma_1&=& - \int  x \Delta_1(x) dx \approx - C_1 \int  \Delta_1(x) dx\nonumber \\& =&C_1 \int  \Delta_2(x) dx \equiv C_1 D_- +C_1 D_+ \label{9}
\end{align}
where we adopted the following definitions:
\begin{align}
D_-&=\int_{x<0}  \Delta_2(x) dx= \dot\gamma\\ 
D_+&=\int_{x>0}  \Delta_2(x) dx  \label{12}
\end{align}
The source $S_2\geq 0$ has support for $x\leq 0$ only, as it corresponds to the flux of sites that become unstable by jumping from the  distribution $P_c(x)$. The positivity of the source implies the positivity of $\Delta_2(x)$. These sites thus start from the unstable region, where they perform a Levy flight and eventually yield. Before doing so, they can jump back to the stable region $x>0$. 
We assume now (as can be checked explicitly in the solutions below) that $\Delta_2(x)$ has significant contributions for $|x|\sim O(1)$ both for positive and negative $x$. This assumption implies that $\int   x \Delta_2(x) dx/D_-$ and $\int   x \Delta_2(x) dx/D_+$ have non-zero limits as the strain rate vanishes, which we write:
\begin{equation}
\delta \sigma_2= - \int   x \Delta_2(x) dx \approx  C_2 D_- -C_3 D_+ \label{13}
\end{equation}
where $C_2, C_3$ are positive constants. Overall we get from Eqs. (\ref{8},\ref{9},\ref{13}):
\begin{align}
\delta \sigma&= (C_1+C_2) D_-  + (C_1-C_3) D_+\nonumber \\
&= (C_1+C_2)\dot \gamma  + (C_1-C_3) D_+
\end{align}

As we shall see below, for $\mu\geq 1$ (which includes the physical case $\mu=1$) we find $ D_{+} \sim \dot{\gamma}^{\alpha}$ with $\alpha<1$, implying that $1/\alpha$ is the Herschel-Bulkley exponent and that $C_1-C_3>0$ (to ensure that $\sigma(\dot\gamma)$ is a growing function as must be the case in this model). For $\mu<1$ we find that the two terms contribute equally and $\beta=1$. To proceed we must thus compute $\alpha$ .

\subsection{ Scaling behavior of $\Delta_2(x)$ for $x<0$}

According to the scaling of $P_c(x)$ described in Eq. (\ref{6}-\ref{7}) and from the definition of $S_2(x)$ in Eq. (\ref{s2}) we obtain $S_2(x)\sim \Theta(-x) |x|^{\theta-\mu}+P_c(0) \delta(x=0)$ for $|x|\ll1$.
 %In the small strain rate limit $\dot{\gamma}\rightarrow 0$, the mechanical noise is small 
Since the average life time of unstable sites is $1$, after jumping in the negative $x$ region, the sites travel a random distance $\Delta x\sim \dot\gamma^H$ as well as a systematic drift of order $\dot \gamma$ before they yield. Here $H$ is the Hurst exponent  $H=1/\mu$ \cite{Bray13}. 

For $\mu\geq 1$, $P_c(0)=0$ and $H\leq1$, thus the fluctuations are always larger than the bias (or as large for $\mu=1$). 
Thus $\Delta_2(x)$ must follow:
 \begin{equation}
 \Delta_2(x)\sim \dot{\gamma}|x|^{\theta-\mu}\label{delta2_neg}
 \end{equation} 
 for $-1\ll x\ll - \Delta x$.  For $|x|\sim \Delta x$ the singularity in Eq. (\ref{delta2_neg}) is rounded off by the fluctuations  and thus:
 \begin{equation}
 \Delta_2(x)\sim  \dot\gamma^{\theta/\mu}  \hbox{ for } |x|\sim \dot\gamma^{1/\mu} \label{14bis}
 \end{equation} 
 For $\mu<1$, $\theta=0$, $P_c(0)>0$ and $H>1$, so the bias dominates and we have:
 \begin{equation}
 \Delta_2(x)\sim \dot{\gamma}|x|^{-\mu}\label{15}
 \end{equation} 
 for $-1\ll x\ll- \dot\gamma$ and:
  \begin{equation}
  \Delta_2(x)\sim 1 \hbox{ for } - \dot\gamma<x<0 \label{15bis}
   \end{equation} 
due to the delta function flux at $x=0$. 
%with an exponential decay at more negative $x$ 
\subsection{ Scaling behavior of $\Delta_2(x)$ for $x>0$}
Ultimately our goal is to compute $\Delta_2(x)$ for $x>0$.  Unstable sites with $x<0$ can escape toward the stable region $x>0$ in two ways. (i) By performing small jumps around $x\sim 0$. This effect turns out to be dominant for $\mu>1$, marginal at $\mu=1$ and negligible for $\mu<1$. It can be taken into account  by imposing the boundary condition Eq. (\ref{14bis}) to the solutions for $\Delta_2(x)$ for $x>0$ discussed below. (ii) By doing large jumps from the negative to positive region, leading to a term $I_1$ for positive $x$ with:
\begin{equation}
I_1=-\int_{-\infty}^{0}A\frac{\Delta_2(y)}{|x-y|^{\mu+1}}dy\label{I1},
\end{equation}
%{\blue We emphasize this source term corresponds to the flux from the negative region to the positive region, in contrast to the source term $S_2$ defined in Eq. (\ref{s2}), which corresponds to the flux from the positive to the negative region due to large jumps and drift.} 
This  term turns out to be negligible for $\mu>1$, marginal for $\mu=1$ and dominant for $\mu<1$, as we now discuss.
 
 \subsection{Case $1<\mu<2$}
For $x>0$, the equation of $\Delta_2(x)$, Eq. (\ref{Deltax}) reduces to: 
 \begin{align}
& \hat{\mathcal{L}}\Delta_2(x)\equiv v_c \frac{d\Delta_2(x)}{dx}  + \int_{0}^{\infty} A\frac{\Delta_2(y)-\Delta_2(x)}{|y-x|^{\mu+1}}dy \nonumber \\
& -\Delta_2(x)\int_{-\infty}^{0}\frac{A}{|x-y|^{\mu+1}}dy = I_1.\label{delta2postive}
 \end{align}
Here, the linear operator $\hat{\mathcal{L}}$ describes a biased Levy Flight with an absorbing condition in $x<0$. We first neglect the term $I_1$ and show that it is indeed self-consistent.
Note that Eq. (\ref{delta2postive}) is very similar to Eq. (\ref{fp2}) describing $P_c(x)$, and can be analized as in \cite{Lin16}. Seeking power-law solutions in $x$ leads to two homogeneous solutions (without the $I_1$ term) for Eq. (\ref{delta2postive}) of the form $\Delta_2(x)= C_3(\dot{\gamma})x^{\mu/2-1}+C_4(\dot{\gamma})x^{\mu/2}$. Because $\Delta_2(x)$ must decay to zero for $x\gtrsim1$, $C_4=0$. Matching the boundary condition Eq. (\ref{14bis}) then implies $C_3(\dot{\gamma})\sim \dot\gamma^{1/\mu}$ and we obtain
\begin{equation}
\Delta_2(x)\sim \dot{\gamma}^{1/\mu} x^{\mu/2-1}\label{delta2_1_pos}.
\end{equation}
After integration we get $D_+\sim \dot\gamma^{1/\mu}$ and therefore 
\begin{equation}
\dot{\gamma}\sim \delta\sigma^{\mu}.
\end{equation}
It is then straightforward to show that  $I_1\sim \dot\gamma $  which is thus  negligible in Eq. (\ref{delta2postive}) where all other terms scale as $\dot\gamma^{1/\mu}$.

%, so we obtain
%\begin{equation}
%\Delta_2(x)\sim\dot{\gamma}^{\alpha}x^{\mu/2-1}.
%\end{equation} 
%In this case, $\Delta_2(x)$ must be continuous on the both sides of $x=0$ with a characteristic length scale $x_c$. Physically Eq. (\ref{Deltax}) represents a group of particles implementing a biased Levy flight towards negative $x$ with drift velocity $-v_c$. The corresponding Hurst exponent $H=1/\mu$, such that the displacement fluctuation after a small strain increment $\gamma$ is $\langle |\delta x|\rangle \sim \gamma^{1/\mu}$. Because the typical time scale $\tau_c$ to yield is $1$, the corresponding strain scale is proportional to the strain rate $\dot{\gamma}$. Since it is the random fluctuation dominates against drift, one can obtain that $x_c\sim \dot{\gamma}^{1/\mu}$. 
%
%To connect $\Delta_2(x)$ on the positive and negative sides, we have $\dot{\gamma} \dot{\gamma}^{\frac{\theta-\mu}{\mu}} \sim \dot{\gamma}^{\alpha} \dot{\gamma}^{\frac{\mu/2-1}{\mu}}$. Using the fact $\theta=\mu/2$, we find $\alpha=1/\mu$. So $D_{+}\sim \dot{\gamma}^{1/\mu}$ and
%\begin{equation}
% \delta\sigma\sim \dot{\gamma}^{1/\mu}.
%\end{equation} 
%The corresponding Herschel-Bulkley exponent $n=1/\mu$. Because $\Delta_2(x)\sim \dot{\gamma}$ for $x<0$, so $I_1\sim \dot{\gamma}$, which confirms the validity of neglecting $I_1$. 

%We can summarize the results as
%\begin{align}
%\Delta_2(x)&\sim \dot{\gamma} |x|^{-\frac{\mu}{2}} \quad\quad\quad -1\ll x \ll -x_c \\
%\Delta_2(x)&\sim \dot{\gamma} x^{\frac{\mu}{2}-1} \quad\quad\quad x_c\ll x \ll 1,
%\end{align}
%with $x_c\sim \dot{\gamma}^{1/\mu}$.
\begin{figure*}[bht!]
 \includegraphics[width=0.95\textwidth]{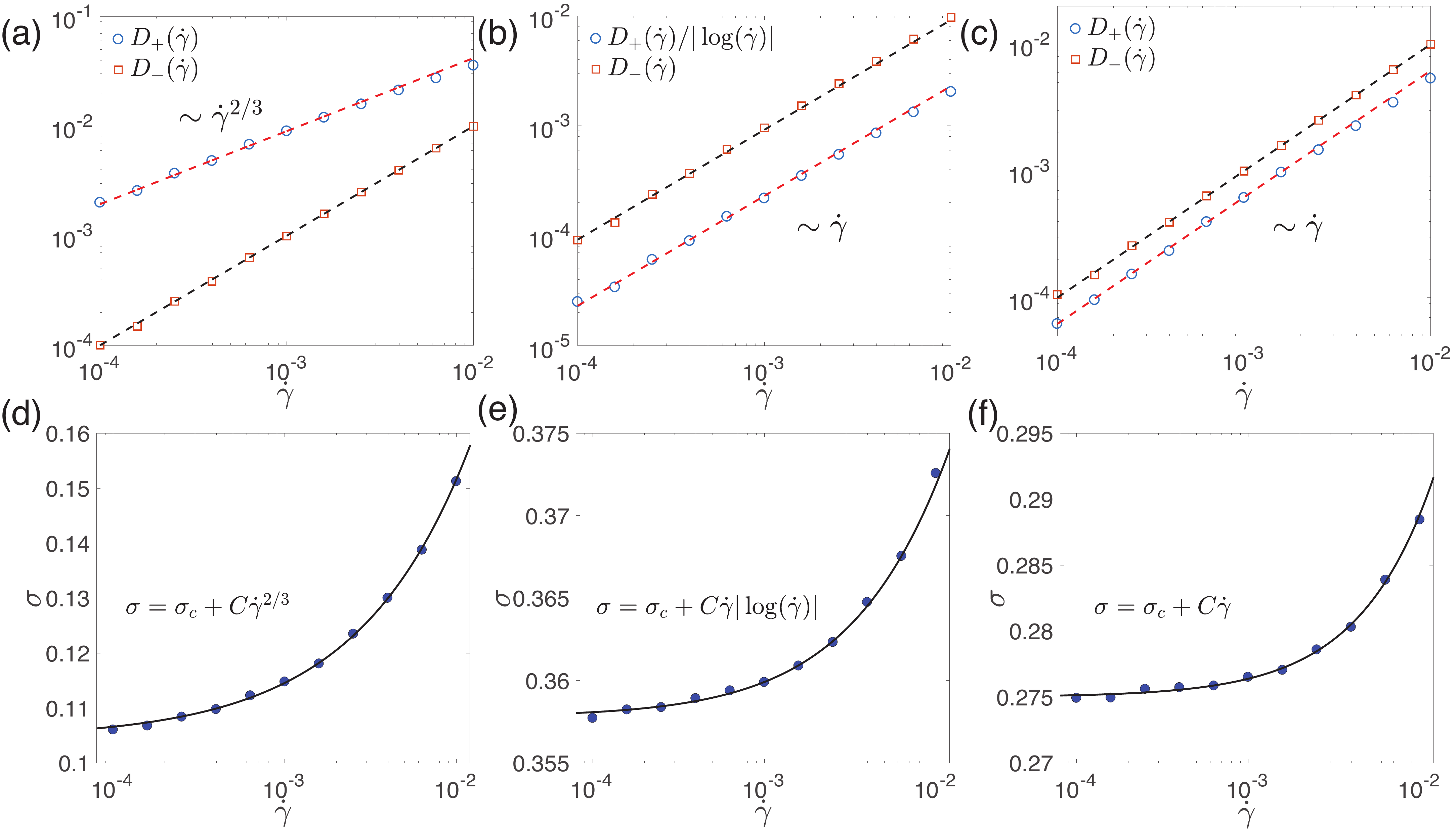}
\caption{(a,b,c)  $D_{+}$ and $D_{-}$ {\it vs} $\dot{\gamma}$. Here, $\mu=1.5$ for (a,d), $\mu=1.0$ for (b,e), and $\mu=0.5$ for (c,f). In all simulations, we take $N=256^2$, and $A=0.3$. We note that to verify the logarithmic correction at $\mu=1$, we plot $D_{+}/|\log(\dot{\gamma})|$ {\it v.s.} $\dot{\gamma}$. (d,e,f) Comparison of the numerical flow curves (blue circles) and the theoretical predictions (lines). }\label{fig2}
\end{figure*}
\subsection{Case $\mu=1$}
This is the physical case.  It is straightforward to show that $I_1\sim \dot{\gamma}x^{\theta-2}$ for $x\ll1$, which is convenient  to write as $I_1= C_5 \dot{\gamma}x^{\delta}$, where the limit  $\delta\rightarrow \theta-2$ will be taken later on. Eq. (\ref{delta2postive}) now becomes
\begin{equation}
\hat{\mathcal{L}}\Delta_2(x)=- C_5\dot{\gamma}x^{\delta}\label{mu1},
\end{equation}
We seek an inhomogeneous solution of the form $B\dot{\gamma}x^{\delta+1}$, implying:
\begin{equation}
(v_cB(\delta+1)-BA+C_5)x^{\delta}=-Bx^{\delta}\int_0^{\infty}A\frac{s^{\delta+1}-x^{\delta+1}}{|s-1|^{2}}dy.
\end{equation}
Using the identity $\int_0^{\infty}\frac{s^{a}-1}{(s-1)^2}ds=1-\pi a\cot(\pi a)$ and $\cot(x)=\cot(x+\pi)$, we obtain:
\begin{equation}
B=\frac{C_5}{(\delta+1)\pi A\{\cot(\pi(\delta+2)-\cot(\pi \theta) \}}\rightarrow  \frac{\tilde{C}_5}{\delta+2-\theta},\label{BB}
\end{equation}
where the limit $\delta\rightarrow\theta-2$ is used. The homogeneous solution of Eq. (\ref{mu1}) that does not grow for large $x$ follows $\Delta_{2h}=C_6 x^{\theta-1}$. Thus  the general solution can be written as:
\begin{equation}
\Delta_2(x) = C_6 x^{\theta-1}+\tilde{C}_5\frac{\dot{\gamma} x^{\delta+1}}{\delta+2-\theta},
\end{equation}
Imposing the boundary condition Eq. (\ref{14bis}) leads to $C_6\sim \dot{\gamma}- \tilde{C}_5\dot{\gamma}^{\delta+3-\theta}/(\delta+2-\theta)$,
implying:
\begin{equation}
\Delta_2(x)=\dot{\gamma}x^{\theta-1}-\tilde{C}_5\frac{\dot{\gamma}^{\delta+3-\theta}}{\delta+2-\theta}x^{\theta-1}+\tilde{C}_5\frac{\dot{\gamma}}{\delta+2-\theta}x^{\delta+1}.
\end{equation}
Taking the limit $\delta\rightarrow 2-\theta$ we obtain:
\begin{equation}
\Delta_2(x)\sim \dot{\gamma}|\log\dot{\gamma}|x^{\theta-1}\label{delta2_2_pos},
\end{equation}
implying that $D_{+}\sim \dot{\gamma}|\log{\dot{\gamma}}|$ and:
\begin{equation}
 \dot{\gamma}\sim \delta\sigma/ |\log{\dot{\gamma}}|,
\end{equation}
showing that the HB exponent is unity with a logarithmic correction.

%Again we can summarize the results as
%\begin{align}
%\Delta_2(x)&\sim \dot{\gamma} |x|^{\theta-1} \quad\quad\quad -1\ll x \ll -x_c \\
%\Delta_2(x)&\sim \dot{\gamma}|\ln\dot{\gamma}| x^{\theta-1} \quad\quad x_c\ll x \ll 1,
%\end{align}
%with $x_c\sim \dot{\gamma}$.
\begin{figure*}[htb!]
\includegraphics[width=0.95\textwidth]{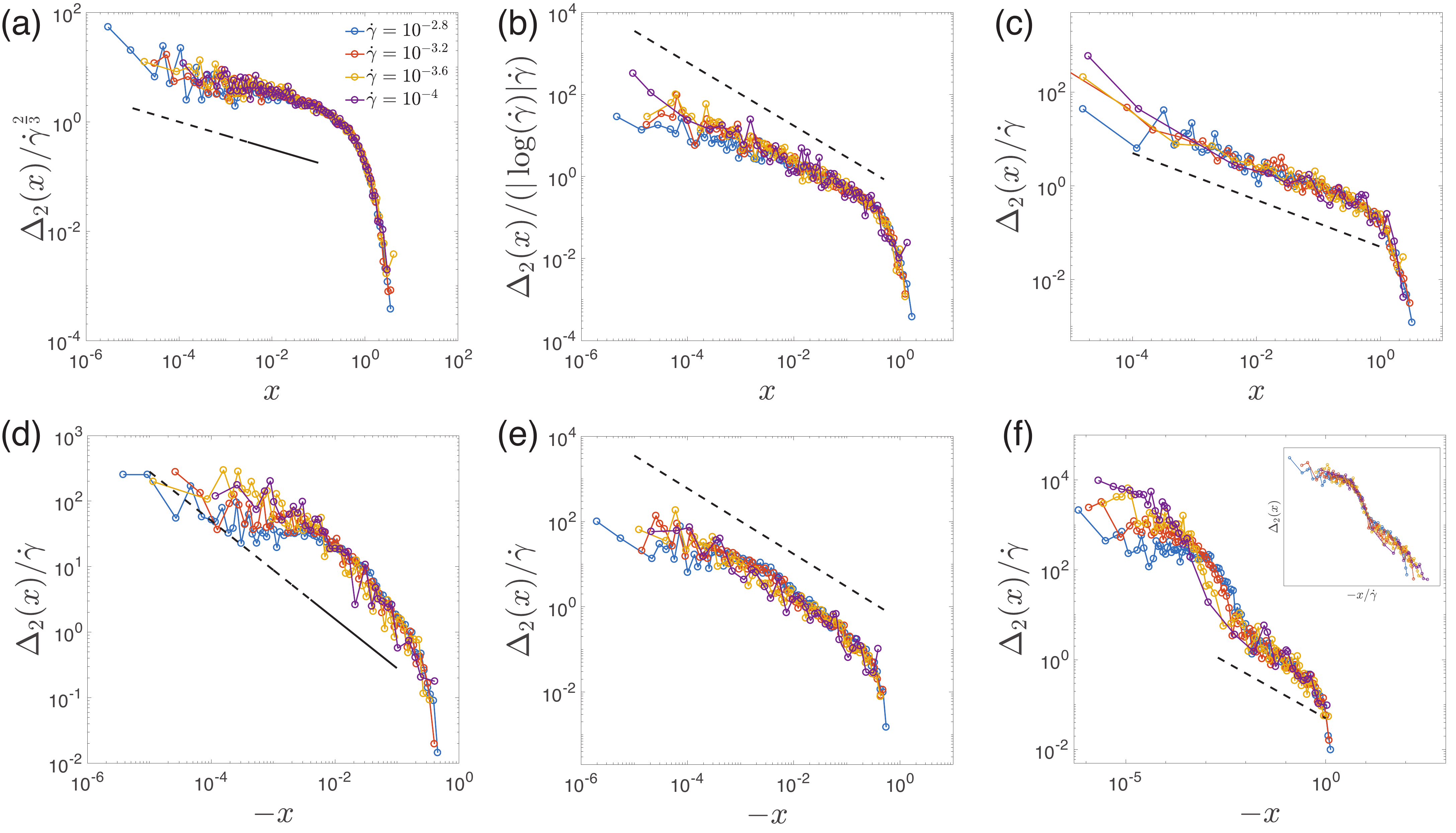}
\caption{(a,b,c) Tests of the predicted scaling collapses of $\Delta_2(x)$ ($x>0$) for $\mu=1.5$ (a), $\mu=1$ (b) and $\mu=0.5$ (c). Colors represent different strain rate, shown in the legend of (a). Dashed lines indicate the slopes predicted by theory. (d,e,f) Collapses of $\Delta_2(x)$ ($x<0$) for $\mu=1.5$ (d), $\mu=1$ (e) and $\mu=0.5$ (f). In the inset of (f), we rescale the $x$ axis to confirm the prediction that the width of the plateau scales as $\dot{\gamma}$ for $\mu<1$.}\label{fig3}
\end{figure*}
\subsection{Case $0<\mu<1$}

In that case the integral defining $I_1$, Eq. (\ref{I1}), is dominated by $-\dot\gamma<x<0$ whose behavior is described in Eq. (\ref{15bis}) and  $I_1\sim\dot{\gamma} x^{-\mu-1}$. Eq. (\ref{delta2postive})  becomes:
\begin{align}
\hat{\mathcal{L}}\Delta_2(x)\sim -\dot{\gamma} x^{-\mu-1}.
\end{align}
No decreasing power-law homogeneous solutions of Eq. (\ref{delta2postive}) can be found in this case,
%$\Delta_2(x)\sim x^{\phi}$ with $-1<\phi<0$ for $x>0$,
%\begin{equation}
%v_c\phi x^{\phi-1} + A\int_{0}^{\infty} \frac{y^{\phi}-x^{\phi}}{|y-x|^{\mu+1}}dy-A\frac{x^{\phi-\mu}}{\mu}=0.\label{phi}
%\end{equation}
%It turns out that both of the second and third term scale as $x^{\phi-\mu}$, which is unable to balance the first term. So $\Delta_2(x)$ is only 
and the solution is only composed of the inhomogeneous one:
\begin{equation}
\Delta_2(x)\sim \dot{\gamma}x^{-\mu}.\label{delta2_3_pos}
\end{equation}
leading to $D_{+}\sim \dot{\gamma}$ and 
\begin{equation}
\dot{\gamma}\sim\delta\sigma.
\end{equation}

Results are summarized in Table \ref{table1}.

\begin{table}[!htb]
\caption{\small{Flow curves in the mean field model, for different Levy coefficient $\mu$, corresponding to a power law interaction exponent $\alpha=d/\mu$.}}\label{table1}
\begin{tabular}{ l ||  r}
\hline\hline
Levy coefficient            &  Flow Curve \\ [3pt] \hline
  $ \mu\geq 2$                      &   $\sigma=\sigma_c+C\dot{\gamma}^{1/2}$      \\[5pt] \hline
  $ 1<\mu<2$                        &    $\sigma=\sigma_c+C\dot{\gamma}^{1/\mu}$     \\[5pt] \hline
  $\mu=1$                            & $\sigma=\sigma_c+C\dot{\gamma}|\log(\dot{\gamma})|$    \\[5pt] \hline
  $\mu<1$                            &   $\sigma=\sigma_c+C\dot{\gamma}$   \\[3pt] \hline  \hline
\end{tabular}
\end{table}

%We first assume the solution is dominated by the last term on the right side from which we obtain $\Delta_2(x)\sim \dot{\gamma}x^{-\mu}$. Given the solution, the first two term scale as $\dot{\gamma}x^{-1-\mu}$ and $\dot{\gamma}x^{-2\mu}$ respectively, not larger than the contribution of the last term, which is consistent with our assumption. The HB exponent in this case becomes $n=1$,

%and we summarize the results as
%\begin{align*}
%\Delta_2(x)&\sim \dot{\gamma} |x|^{-\mu} \quad\quad\quad -1\ll x \ll -x_c \\
%\Delta_2(x)&\sim \dot{\gamma}x^{-\mu} \quad\quad\quad\quad x_c\ll x \ll 1,
%\end{align*}
%and $x_c\sim \dot{\gamma}$ since the Levy flight is dominated by the drift in this case.
\section{\label{numerics}Numerical Simulations}
 We use simulation to test our theoretical results. In Eq. (\ref{dynamics}), the shear stress is fixed. This is equivalent to fixing the strain rate in the thermodynamic limit. We do the latter, as it minimizes finite size effects. We simulate a modified version of Eq. (\ref{dynamics}), $\sigma_j(l+1)-\sigma_j(l)= \delta\sigma_0(l)+\delta \sigma_j $, where $\delta\sigma_0(l)$ is adjusted at each step to maintain a constant fraction of unstable sites, which is the strain rate $\dot{\gamma}$. We compute $\Delta_2(x)$ as those sites which entered the unstable region $x<0$ at least once during their life time. After obtaining $\Delta_2(x)$, we compute $D_{+}$ and $D_{-}$ as its integral in the positive and negative region. 

%  Numerically, it is more convenient to keep the strain rate $\dot{\gamma}$ constant for finite systems. To achieve this, we slightly modify the rule and adjust the stress level to maintain a constant fraction of unstable sites,
%\begin{align}
%\sigma_j(l+1)-\sigma_j(l)=\delta\sigma(l)+\delta\sigma_j,\\
%\delta\sigma(l)=1-\sigma_m(l)+\frac{\sigma_m(l)-\sigma_{m+1}(l)}{2}
%\end{align}
%where $\sigma_m$ is the site which carries the $m_{th}$ largest local stress and $\delta\sigma$ is chosen to maintain the number of unstable sites as $m=N\dot{\gamma}$ after local yielding.
In the upper row of Fig. \ref{fig2}, we plot the scaling of $D_{+}$, $D_{-}$ against $\dot{\gamma}$ for three representing cases $\mu=3/2$, $\mu=1$, and $\mu=1/2$. For all cases we get $D_{-}=\dot{\gamma}$ and different scaling of $D_{+}$. Our theoretical predictions for $D_{+}$, indicated as dashed lines in the figure,  are nicely verified. In the bottom row, we plot the flow curves with the theoretical predictions where the HB exponent is fixed and the critical stress $\sigma_c$ and coefficient $C$ is fitted, showing again an excellent match with the theoretical predictions.

In Fig. \ref{fig3}, we test the scaling of $\Delta_2(x)$ for the same $\mu$ by collapsing the data according to our theoretical predictions without any fitting parameters. In the upper row, we test the scaling of $\Delta_2(x)$ for $x>0$, Eq. (\ref{delta2_1_pos}, \ref{delta2_2_pos}, \ref{delta2_3_pos}) against the predicted power law exponent shown as dashed lines, and get satisfying agreement. In the bottom row, we test Eq. (\ref{delta2_neg}) for $x<0$. Finally,  in the inset of Fig. \ref{fig3}f, we rescale the $x$ axis by $\dot{\gamma}$ to test Eq. (\ref{15bis}).

\section{\label{finitedimension} Comparison with finite-dimensional elasto-plastic models. }

We have argued that the mean-field model with proper noise statistics  leads to $\beta=1$  (with logarithmic corrections that can suggest a slightly larger exponent if fits are restricted on a small range of strain rates, say  1.2 instead of one for the typical dynamical range studied in the literature). In elasto-plastic models in finite dimensions, we found $\beta\approx 1.52$ and $\beta\approx1.38$ in two and three dimensions respectively \cite{Lin14}. Our mean field result $\beta=1$ (with a logarithmic correction) is thus consistent with  the observation that  $\beta$ decreases as the dimension increases, and is not too far off from the $d=3$ value (especially considering the effect of the logarithm). It would be interesting to measure the exponent $\beta$ in $d=4$, where the MF predictions for the pseudo-gap exponent $\theta$ appear to become exact \cite{Lin16}. 

\section{\label{finitedimension} Finite propagation speed of elastic information }

A limitation of elasto-plastic models is that they tend to give values for $\beta$ smaller than those observed in molecular-dynamics simulations and experiments (see e.g. the table in \cite{Lin14}), whereas other exponents they predict (on avalanche statistics and fractal dimension as well as on the pseudo-gap exponent $\theta$) appear consistent with molecular dynamics \cite{Lin14, Liu16, Salerno13,Karmakar10a,Maloney04,Arevalo14,Bailey07}. Several simplifications of these models  could be responsible for this discrepancy. In particular, the elastic coupling is established instantaneously throughout the system (while in fact they should be established ballistically $r\sim t$ for inertial systems and diffusively $r\sim\sqrt{t}$ for over-damped ones). We believe that this effect is responsible for the difference  in the dynamics between elasto-plastic models and MD simulations. %{\blue In our previous work \cite{Lin14a}, we constructed a scaling description for finite dimensional systems corroborated with simulations of elasto-plastic model in two and three dimensions based on instantaneous elastic coupling. We investigated the flow curves and found $\beta\approx 1.52$ and $\beta\approx1.38$ in two and three dimensions respectively. Our mean field result $\beta=1$ (with a logarithmic correction) is consistent with \cite{Lin14a} in which $\beta$ decreases as the dimension increases. 

In the following we extend the scaling description developed in \cite{Lin14} to incorporate the fact that the elastic signal has a finite speed to travel and show that this modification leads to larger $\beta$ values, closer to experiments and molecular dynamics simulations \cite{Salerno12,Chaudhuri12,Mobius10,Becu06}. We first recall some aspects of the scaling description of the yielding transition introduced in \cite{Lin14}. We define two exponents, respectively the fractal dimension and the dynamic exponent of avalanches:
\begin{align}
S_c&\sim L^{d_f},  \\
T_c&\sim L^{z}.
\end{align}
Here $S_c$, $T_c$ are the characteristic avalanche size and duration of a finite system size $L$. The hyper-scaling scaling relation:
\begin{equation}
\beta=1+\frac{z}{d-d_f}
\end{equation}
has been shown to hold for $d=2$ and $d=3$ in  elasto-plastic models \cite{Lin14,Liu16} and other models where the finite speed of interactions is included \cite{Jagla17} (it is currently unclear if there exists an upper critical dimension $d_c$ beyond which this  relation breaks down).  However, elasto-plastic models find $z<1$  which cannot be true asymptotically in real materials, since information cannot propagate faster than ballistically, implying $z\geq 1$.  As is the case for the depinning transition with long-range interaction, the observation that models with instantaneous kernel find $z<1$ supports that the above constraint is saturated, leading to $z=1$ \cite{Fisher97}. Thus we expect that  asymptotically:
\begin{equation}
\beta=1+\frac{1}{d-d_f}
\end{equation}
Our previous elasto-plastic measurements \cite{Lin14} in two and three dimensions then suggest $\beta\approx2.1$ (because $d_f\approx1.1$ for $d=2$) and $\beta\approx 1.7$ ($d_f\approx1.5$ for $d=3$).  The underlying assumption in these numerical estimates is that the fractal dimension of avalanches is not affected by the choice of dynamics. This is consistent with observations, but is not proven (unlike for the depinning transition where the monotonicity of the interaction implies that the avalanche statics cannot depend on the dynamical rules).

%
%
% Apart from the difference in $z$, elasto-plastic models turn out to be quantitatively correct in predicting $d_f$, $\theta$ and $\tau$ \cite{Lin14,Liu16}. We can thus give effective predictions using the relation expressing $\beta$ in terms of $z$ (assuming $z=1$) and $d_f$ from our previous elasto-plastic measurements in two and three dimensions and obtain $\beta\approx2.1$ ($d_f\approx1.1$, 2D) and $\beta\approx 1.7$ ($d_f\approx1.5$, 3D), larger than what we reported previously, closer to molecular dynamics simulations and experiments \cite{Karmakar10,Chaudhuri12,Cloitre2003,Mobius10}. 
%
%

It is useful to express this result in terms of the exponent $\tau$, that characterizes the power law distribution of the avalanche size distribution $P(S)\sim S^{-\tau}$. Using the scaling relation  $\tau=2-\frac{\theta}{\theta+1}\frac{d}{d_f}$ \cite{Lin14} one gets:
\begin{equation}
\beta=1+\frac{(1+\theta)(2-\tau)}{d(1+\theta)(2-\tau)-\theta}
\end{equation}
Making the additional approximation that $\tau\approx 3/2$ (as is often the case for mean-field crackling systems, and seems to be already a pretty good approximation for $d=3$ where $\tau\approx 1.45$ \cite{Lin14} but less so for $d=2$) leads to:
\begin{equation}
\beta\approx1+\frac{1+\theta}{d(1-\theta)}.\label{betatheta}
\end{equation}
This expression relates $\beta$ to the pseudo-gap exponent $\theta$, which can be computed with the present mean-field approach. Observations indicate $\theta\approx 0.35 $ for $d=3$ and $\theta_{MF}\approx 0.3$ \cite{Lin16}. Thus from this result and Eq. (\ref{betatheta}), we obtain an approximate analytic prediction for $\beta\approx 1.6$ for $d=3$ (and obtain again $\beta\approx 1.7$ if the observed value fo $\theta$ is injected in Eq. (\ref{betatheta})).

\section{Discussion: role of potential smoothness.}

{\bf Mean-field:} A second simplification of elasto-plastic models is that the time scale for a site to become unstable $\tau_c$ is assumed to be independent of how unstable the site is. However, very weakly unstable sites with $0<-x\ll1$ should take a  longer time to relax, an effect that could be incorporated by allowing for a $x$-dependent evaporation rate $\tau(x)$. This phenomenon plays a role in   mean field depinning models, as shown by Fisher \cite{Fisher85}. In this limit, an elastic manifold is represented by a collection of sites, all coupled to each other. Each site lies in a disordered potential. In such models, it is found that the exponent $\beta$ differs if the potential is smooth or if it present cusps. This difference comes from the fact that for smooth potential, a barely unstable site spends a lot of time running down its potential, which does not occur for cuspy potential. However, this distinction disappears in finite dimensions, because in that case sites that become unstable  do so by receiving a big kick from the rearrangement of a neighboring site, so that the condition $0<-x\ll1$ is typically not satisfied (in mean-field depinning it is satisfied, because kicks are small). In this view, we argue that this effect should not play any role in a proper mean-field description of plasticity.  Indeed when the fat tail of mechanical noise is considered,  unstable sites typically became so after receiving a stress kicks of order $1$ to cross the threshold (this statement is equivalent to the fact that the integral of Eq. (\ref{delta2_neg}) is not dominated by small $x$), just like for depinning in finite dimension. From this argument, we expect that effects associated with the smoothness of the potential will play a role only in less realistic mean-field models of the yielding transition in which the mechanical noise does not display fat tails (such as in the Hebraud-Lequeux model), since in that case  sites that become unstable always do so at $x=0$.  The recent numerical work of Jagla \cite{Jagla17} on a variation of the mean-field model of HL in which sites are described by a random potential (see Table 1 of \cite{Jagla17}) is an illustration of our views:
in that case smoothness matters, an artefact of the choice of mechanical noise used. 

%Therefore, our mean field model, which we believe is more physical, is not sensitive to the shape of the strain energy.

 {\bf Finite dimension:} 
% The recent numerical work of Jagla \cite{Jagla17}, in which each site is characterized by a disordered potential energy shows that the exponent $\beta$ is non-universal and differs if the disorder is smooth (in which case $\tau(x)$ diverges as $x\rightarrow 0$) or presents cusps (where assuming that $\tau(x)$ is constant is justified). In the mean field level of Jagla's model, the interaction between local events are Gaussian distributed, with the interaction strength scale inversely with the square root of system size. One could expect the different universality of HB exponent in these cases, because those unstable sites typically cross the local yield strain due to infinitely small fluctuations. Therefore, the detailed shape of potential energy is crucial to the relaxation dynamics of yielding sites and for the smooth potential, it will take a much longer time to relax compared with the cusp potential. Similar phenomena are also observed in the mean field depinning model, as shown by Fisher \cite{Fisher85}. In fact, one version of his mean field model (see Table 1 of \cite{Jagla17}) is essentially the same as Hebraud-Lequeux model \cite{Hebraud98}. While in the correct mean field where the fat tail of stress fluctuation is considered, those unstable sites typically receive stress kicks of order $1$ to cross the threshold as we have shown (since the integral of Eq. (\ref{delta2_neg}) is not dominated by small $x$). Therefore, our mean field model, which we believe is more physical, is not sensitive to the shape of the strain energy.
% 
 The smoothness of the potential may however affect the dynamics in finite dimensions, because sites that become unstable due to a rearrangement at a large distance will have $0<-x\ll1$. Although such spatially extended  jumps of activity are not the typical ones, they may play a special role for the dynamics of avalanches. This view is supported by the finite dimensional model of Jagla \cite{Jagla17} which does find that the scaling exponents near the yielding transition, including the exponents $d_f$ and $\tau$ defined above, depend on the smoothness of the site potential.   However, there are several indications that in that work the thermodynamic limit is not probed. It is found that the avalanche size exponent $\tau$ is smaller than $1$, which is impossible asymptotically since $\tau$ characterizes the distribution of avalanche size. Moreover, the dynamic exponent $z$ is found to be smaller than one, which violates the limited speed of information in elastic materials. Since finite size effects are known to lead to a spurious dependence of the dynamics on the nature of the disorder \cite{Myers93}, it is currently unclear if this non-universality persists or not near the yielding transition in the thermodynamic limit. More work is needed to clarify these points.

\section{Conclusion}
 In this work, we have computed the HB exponent within a mean-field approximation that includes the fat tails characterizing the mechanical noise. We obtain $\beta=1$, a value quite smaller than a previous mean-field calculation where noise was assumed to be Gaussian for which $\beta=2$ \cite{Hebraud98}, and also smaller than empirical measurements \cite{Salerno12,Chaudhuri12,Mobius10,Becu06}. This result indicates that finite-dimensional effects must be included to get satisfying results for the dynamics (whereas for static properties like the pseudo-gap exponent $\theta$, our mean-field approximation is already quite accurate in $d=3$ \cite{Lin15a,Lin14,Salerno13,Karmakar10a}). We have argued based on observations in elasto-plastic models that the dynamic exponent must follow $z=1$, leading to a potentially exact expression of $\beta$ in terms of the fractal dimension of the avalanches $d_f$. Making an approximation on the avalanche statistics leads to an expression of $d_f$ or $\beta$ in terms of the pseudo-gap exponent $\theta$, which can be computed analytically, ultimately yielding a theoretical approximation for $\beta$  in $d=3$. It would be very interesting to obtain accurate measurements of $\beta$ in experiments or MD simulations to test these predictions.  At a theoretical level, questions for the future include the computation of avalanche exponents, in particular $\tau$, in mean-field. Although the usual value $\tau=3/2$ is consistent with our numerics (we find $\tau\approx 1.45$, not shown) it is yet not derived for the dynamics studied here where unstable sites can evaporate at a finite rate. Ultimately, a  precise computation of $\beta$ would require to compute finite-dimensional corrections to the mean-field presented here. %of a perturbation theory near this dimension to compute systematically corrections to mean-field approximations. 
 
 \section*{ACKNOWLEDGMENTS}
It is a pleasure to thank E. Agoritsas, J.P. Bouchaud, T. De Geus, P. Le Doussal, M Muller, W. Jie and  A. Rosso for discussions related to this work. MW thanks the Swiss National Science Foundation for support under Grant No. 200021-165509 and the Simons Foundation Grant (\#454953 Matthieu Wyart).

\bibliography{Wyartbibnew}
\end{document}